\documentclass[a4paper,12pt]{article}
\usepackage{amsfonts}
\makeatletter
\def\numberbysection{\@addtoreset{equation}{section}
        \def\theequation{\thesection.\arabic{equation}}}
\def\eqnarray{%
   \stepcounter{equation}%
   \def\@currentlabel{\p@equation\theequation}%
   \global\@eqnswtrue
   \m@th
   \global\@eqcnt\z@
   \tabskip\@centering
   \let\\\@eqncr
   $$\everycr{}\halign to\displaywidth\bgroup
       \hskip\@centering$\displaystyle\tabskip\z@skip{##}$\@eqnsel
      &\global\@eqcnt\@ne\hfil$\displaystyle{\hbox{}##\hbox{}}$\hfil
      &\global\@eqcnt\tw@ $\displaystyle{##}$\hfil\tabskip\@centering
      &\global\@eqcnt\thr@@ \hb@xt@\z@\bgroup\hss##\egroup
         \tabskip\z@skip
      \cr
}
\numberbysection
\def\bW {\kern 0.1em\overline{\kern -0.1em W \kern -0.1em} \kern
0.1em}
\def\bcW {\kern 0.1em\overline{\kern -0.1em \mathcal W \kern -0.1em}
\kern 0.1em}
\def\cW{\mathcal W}

\def\mbar#1{\kern 0.1em\overline{\kern -0.1em #1 \kern -0.1em} \kern
0.1em}

\def\d{\mathrm d}
\mathchardef\Delta="101

\mathchardef\Gamma="100
\mathchardef\Lambda="103

\def\g{\mathfrak g}
\def\gl{\mathfrak{g}\mathfrak{l}}

\mathchardef\Phi="108
\def\R{\mathbb R}
\def\tr{\mathrm{tr}}
\mathchardef\Upsilon="107

\def\Z{\mathbb Z}
\makeatother
\begin{document}

\begin{titlepage}
\hbox to \hsize{\hfil {\tt INR-TH-02-10/1}}
\hbox to \hsize{\hfil {\tt October 2002}}
\vfill
\large \bf
\begin{center}
Higher Symmetries of Toda Equations
\end{center}
\vskip10mm
\normalsize
\begin{center}
Khazret S. Nirov${}^a${}\footnote{E-mail: nirov@ms2.inr.ac.ru}
$\,$
and Alexander V. Razumov${}^b${}\footnote{E-mail: razumov@mx.ihep.su}\\
{\small \it ${}^a{}$Institute for Nuclear Research of the Russian
Academy of Sciences,} \\ 
{\small \it 60th October Anniversary Prospect 7a, 117312 Moscow,
Russia}\\
{\small \it ${}^b{}$Institute for High Energy Physics,
142280 Protvino, Moscow Region, Russia}
\end{center}
\vskip20mm
\begin{abstract}
The symmetries of the simplest non-abelian Toda equations are
discussed. The set of characteristic integrals whose Hamiltonian
counterparts form a $W$-algebra, is presented.
\end{abstract}
\vskip40mm
\begin{center}
{\it Based on talk given at 12th International Seminar
on High Energy Physics}\\ 
{\it Quarks'2002, Novgorod, Russia, June 1--7, 2002}
\end{center}
\vfill
\end{titlepage}

%%%%%%%%%%%%%%%%%%%%%%%%%%%%%%%%%%%%%%%%%%%%%%%%%%%%%%%%%%%%%%%%%%%%
\hspace{20mm}
%%%%%%%%%%%%%%%%%%%%%%%%%%%%%%%%%%%%%%%%%%%%%%%%%%%%%%%%%%%%%%%%%%%%

\section{Introduction}

The main way to investigate nonlinear differential equations is
through revealing and analysing their symmetries. The ideological
foundation for this approach was laid at the end of the nineteenth and
the beginning of twentieth centuries by S.~Lie and E.~Noether. Since
then the most interesting and important results in this field of
research have been obtained in development of the theory of integrable
systems; for a quite exhaustive list of literature see
monographs~\cite{Olv,FTa}. 

A principal place among nonlinear integrable equations occupy
two-\-di\-men\-sio\-nal systems which are representable as
the ze\-ro-\-cur\-va\-tu\-re condition for some connection on a
trivial principal fiber bundle \cite{FTa,LSa,RSa}. This 
transparent background allows for the most effective application 
of group-algebraic and differential-geometry methods to the study 
of such systems. A class of that sort, called the non-abelian Toda
equations being a subclass of the Toda-type integrable systems
\cite{RSa}, is considered in the present talk. Actually
here we deal with the simplest examples of non-abelian Toda systems
and concern only the symmetry aspects of the theory.

\section{Non-abelian Toda systems}

Let $G$ be a real or complex matrix Lie group whose Lie algebra $\g$
is endowed with a $\Z$-gradation,
\[
\g = \bigoplus_{m \in \Z} \g_m, \qquad [\g_m,\g_n] \subset \g_{m+n}.
\]
Assume that the $\Z$-gradations under consideration is generated by a
grading operator. It means that there exists an element $q \in \g$,
called the grading operator, such that
\[
\g_m = \{x \in \g \mid [q, x] = m x\}.
\]
It is known that any $\Z$-gradation of a semisimple Lie algebra is
generated by a grading operator.

{}From the definition of a $\Z$-gradation it follows that the subspace
$\g_0$ is a Lie subalgebra of $\g$. Denote by $G_0$ the connected Lie
group corresponding to this subalgebra. Let $l$ be a positive integer
such that the grading subspaces $\g_m$ are trivial for $-l < m < 0$
and $0 < m < l$. In accordance with the group-algebraic approach
\cite{LSa, RSa}, the Toda equations are defined as the matrix
differential equations
\begin{equation}
\partial_+ \left( \gamma^{-1} \partial_- \gamma \right)
= [ c_-,\gamma^{-1} c_+ \gamma ].
\label{TE}
\end{equation}
Here $\gamma$ is a mapping from the manifold $\R^2$ with the
coordinates denoted by $x^-$ and $x^+$, $c_-$ and $c_+$ are some fixed
mappings from $\R^2$ to $\g_{-l}$ and  $\g_{+l}$, respectively,
satisfying the conditions
\[
\partial_+ c_- = 0, \qquad \partial_- c_+ = 0.
\] 
As usually, we denote the partial derivatives over $x^-$ and $x^+$ by
$\partial_-$ and $\partial_+$. There exist the so-called higher
grading \cite{Lez95,GeSa} and multi-dimensional \cite{RSa97a, RSa97b}
generalisations of the Toda systems.

If the $\Z$-gradation under consideration is such that the resulting
Lie subalgebra $\g_{0}$ and therefore the corresponding Lie subgroup
$G_0$ are abelian, then the corresponding Toda system is called
abelian. In any other case one has a non-abelian Toda system. There is
a complete classification of the Toda systems associated with
classical complex Lie groups \cite{RSaZu}.

The Toda equation (\ref{TE}) can be derived from the action 
functional \cite{GeSa}
\[
S[\gamma] = S_{\mathrm{WZNW}}[\gamma] + S_{\mathrm T}[\gamma],
\]
where $S_{\mathrm{WZNW}}[\gamma]$ is the action functional for the
Wess--Zumino--Novikov--Witten (WZNW) model on the Lie group $G_{0}$
\cite{Nov, Wit}, and $S_{\mathrm T}[\gamma]$ is a no-derivatives term 
\[
S_{\mathrm T}[\gamma] = \kappa \int \d x^- \d x^+ \,
\mathrm{Tr} \left( \gamma^{-1} c_+ \gamma c_- \right).
\]
Here $\kappa$ is a free parameter entering the action of the WZNW
model, and Tr~is the appropriately normalized trace.

As a concrete example we consider in this talk some class of
non-abelian Toda systems associated with the real general linear Lie
group $\mathrm{GL}_{\,2n}(\R)$ defined as follows. Let the Lie algebra
$\gl_{\,2n}(\R)$ is supplied with the $\Z$-gradation generated by the
grading operator
\[
q = \frac{1}{2}
         \left( \begin{array}{cc}
           I_n & 0 \\
              0    & -I_n
         \end{array} \right),
\]
where $I_n$ is the unit $n \times n$ matrix. In this case $\g =
\g_{-1} \oplus \g_{0} \oplus \g_{+1}$ and $l=1$. Choose the mappings
$c_-$ and $c_+$ as
\[ 
c_- = 
         \left( \begin{array}{cc}
           0     & 0 \\
           I_n   & 0
         \end{array} \right),
\qquad
c_+ = 
         \left( \begin{array}{cc}
           0   & I_n \\
           0   & 0
         \end{array} \right).
\] 
The Lie group $G_{0}$ turns out to be isomorphic to the direct product
of two copies of the general linear Lie group $\mathrm{GL}_n(\R)$, and  
the mapping $\gamma$ has the form
\[ 
\gamma = 
         \left( \begin{array}{cc}
           \Gamma^{(1)} & 0 \\
              0    & \Gamma^{(2)}
         \end{array} \right), 
\]  
where $\Gamma^{(1)}$ and $\Gamma^{(2)}$ are mappings from $\R^2$ to
$\mathrm{GL}_n(\R)$. Toda equation (\ref{TE}) decomposes into two
matrix differential equations
\[
\partial_+ \left( \Gamma^{(1)-1} \partial_- \Gamma^{(1)} \right) 
= - \Gamma^{(1)-1} \Gamma^{(2)}, 
\quad
\partial_+ \left( \Gamma^{(2)-1} \partial_- \Gamma^{(2)} \right) 
= \Gamma^{(1)-1} \Gamma^{(2)}. 
\]
The general solution to these equations was constructed in paper
\cite{RaSa97}.

The action functional for the system under consideration is
\[
S[\Gamma^{(1)}, \Gamma^{(2)}] = S_{\mathrm{WZNW}}[\Gamma^{(1)}] +
S_{\mathrm{WZNW}}[\Gamma^{(2)}] + S_{\mathrm{T}}[\Gamma^{(1)},
\Gamma^{(2)}],
\]
where $S_{\mathrm{WZNW}}[\Gamma^{(1)}]$ and
$S_{\mathrm{WZNW}}[\Gamma^{(2)}]$ are the actions for the WZNW model
on the above mentioned two copies of the group $\mathrm{GL}_n(\R)$,
and the no-derivatives term is now of the form
\[
S_{\mathrm{T}}[\Gamma^{(1)}, \Gamma^{(2)}] = \kappa \int \d x^- \d x^+
\, {\tr} \left( \Gamma^{(1)-1} \Gamma^{(2)} \right).
\]
Here tr means the usual trace.

Indubitable advantage given by the formulation of the system in
terms of the action functional is that it allows us to proceed to the
Hamiltonian description, which in turn provides a way to consider the
symmetries of the system as the transformations generated by conserved
charges. The construction of the Hamiltonian formalism can be
performed in the spirit of paper~\cite{Bo}.

\section{Simplest symmetries}

The action $S_{\mathrm{WZNW}}[\gamma]$ is invariant with respect to
the transformation
\begin{equation}
\gamma(x^-, x^+) \to \lambda_+(x^+) \, \gamma(x^-, x^+) \,
\lambda_-^{-1}(x^-), \label{sym1}
\end{equation}
where $\lambda_-$ and $\lambda_+$ are arbitrary $G_0$-valued
functions. The total action $S[\gamma]$ of the Toda system inherits
only a part of this symmetry. Indeed, it is clear that transformation 
(\ref{sym1}) leads again to the action of the Toda system with the
mappings $c_-$ and $c_+$ changed as follows:
\[ 
c_- \to \lambda_- c_- \lambda_-^{-1},
\qquad
c_+ \to \lambda_+ c_+ \lambda_+^{-1}. 
\] 
Hence, if the matrix-valued functions $c_-$ and $c_+$ are such
that
\[ 
\lambda_- c_- \lambda_-^{-1} = c_-, \qquad
\lambda_+ c_+ \lambda_+^{-1} = c_+,
\]  
then (\ref{sym1}) is a symmetry transformation for the Toda system
under consi\-deration. It is worthwhile to call this invariance a
WZNW-type symmetry.

For the concrete example of a non-abelian Toda system introduced in the
previous section the WZNW-type symmetry is realised as
\begin{eqnarray}
& \Gamma^{(1)}(x^-, x^+) \to \Lambda_+(x^+) \Gamma^{(1)}(x^-, x^+)
\Lambda_-^{-1}(x^-), \label{ws1} \\
& \Gamma^{(2)}(x^-, x^+) \to \Lambda_+(x^+) \Gamma^{(2)}(x^-, x^+)
\Lambda_-^{-1}(x^-). \label{ws2}
\end{eqnarray}
It is clear that $S_{\mathrm T}[\gamma]$ is invariant with respect to
these transformations.

The action of the WZNW model is invariant also with respect to the
conformal transformations
\[
x^- \to f^-(x^-), \qquad x^+ \to f^+(x^+)
\]
acting on the mapping $\gamma$ in accordance with the rule
\[
\gamma(x^-, x^+) \to \gamma(f^-(x^-), f^+(x^+)).
\]
The action of a Toda system is not invariant with respect to these
transformations. Nevertheless, in the case where $c_-$ and $c_+$ are
constant mappings, one can demonstrate that the corresponding Toda
system is invariant with respect to the slightly modified action of the
conformal group. To this end, one can first get convinced that if the
mapping $\gamma$ satisfies the equation (\ref{TE}), then the mapping
$\gamma'(x^-, x^+) = \gamma(f^-(x^-), f^+(x^+))$ satisfies the
equation
\[ 
\partial_+ \left( \gamma^{\prime -1} \partial_- \gamma' \right)
= \partial_- f^- \partial_+ f^+ [ c_-, \gamma^{\prime -1} c_+
\gamma'].
\] 
One can compensate the factor $\partial_- f^- \partial_+ f^+$
at the right hand side of this equation using the transformation
(\ref{sym1}). To this end, one should put
\[
\lambda_- = \exp (+q \, l^{-1}\ln \partial_- f^-), \qquad
\lambda_+ = \exp (-q \, l^{-1}\ln \partial_+ f^+),
\]
where $q$ is the grading operator. It leads to the relations
\[
\lambda_- c_- \lambda_-^{-1} = (\partial_-f^-)^{-1} c_-,
\qquad
\lambda_+ c_+ \lambda_+^{-1} = (\partial_+f^+)^{-1} c_+,
\]
allowing exactly for the desirable cancellation of the superfluous
factor. As the result, the conformally transformed mapping
\begin{equation}
\gamma'' = \exp(-q \, l^{-1}\ln \partial_+f^+) \,
\gamma' \exp(-q \, l^{-1}\ln \partial_-f^-) \label{ct}
\end{equation}
satisfies the initial Toda equation. We see that the space of
solutions of the Toda equations is invariant under the appropriately
defined action of the group of conformal transformations \cite{RSaZu}. 

\section{\mathversion{bold}Characteristic integrals and
$W$-symmetries}

The WZNW-type symmetry and the conformal symmetry do not exhaust all
symmetries of a Toda system. To find additional symmetry
transformations we can use the following procedure \cite{NiR}. 

First we find some set of conserved charges. In the case under
consideration we have an infinite set of conserved charges provided
by the so-called characteristic integrals. In the Hamiltonian
formalism conserved charges generate symmetry transformations. So, we
construct the Lagrangian formulation for our system and then proceed
to the corresponding Hamiltonian formulation. After that we consider
the symmetry transformations generated by the Hamiltonian counterparts
of the conserved charges associated with the characteristic integrals,
and finally obtain their Lagrangian version. This allows us, in
particular, to obtain the WZNW-type symmetry transformations and the
conformal transformations discussed above.

A characteristic integral of a Toda system is, by definition, either
a differential polynomial $W$ of the Toda fields satisfying the
relation
\begin{equation}
\partial_+ W = 0, \label{ci1}
\end{equation}
or a differential polynomial $\bW$ of the Toda fields which satisfy
the relation
\begin{equation}
\partial_- \bW = 0. \label{ci2}
\end{equation}
By a differential polynomial we mean a polynomial function of the
fields and their derivatives. The existence of the characteristic 
integrals, under appropriate conditions, guarantees the integrability
of the Toda equations~\cite{LSSh+}.

Let us treat the manifold $\R^2$ as a flat Riemannian manifold with
the coordinates $x^-$ and $x^+$ being light-front coordinates. The
usual flat coordinates $x^0 = t$ and $x^1 = x$ are related to the
light-front coordinates by the relation
\[
x^0 = x^- + x^+, \qquad x^1 = {} - x^- + x^+.
\]
Using these coordinates, we write equality (\ref{ci1}) as
\[
\partial_t W + \partial_x W = 0,
\]
where $\partial_t = \partial/\partial t$ and $\partial_x =
\partial/\partial x$. Hence, the function $W$ is a density of a
conserved charge. Moreover, multiplying $W$ by a function which
depends only on $x^-$ we again obtain a characteristic integral.
Therefore, a characteristic integral generates an infinite set of
densities of conserved charges. Similarly, multiplying a
characteristic integral satisfying relation (\ref{ci2}) by functions
depending only on $x^+$ we again obtain an infinite set of densities
of conserved charges.

It is clear that any differential polynomial of characteristic
integrals is also a characteristic integral. Moreover, the Poisson
bracket of the Hamiltonian counterparts of any two characteristic
integrals is again a characteristic integral. Therefore, a necessary
step in investigation of characteristic integrals is to show that
they form a closed set with respect to the Poisson bracket, or, in
other words, that they form an object called a $W$-algebra. The
elements of a $W$-algebra generate the so-called $W$-symmetries being
a nonlinear extension of the conformal symmetry. The study of such
extensions was initiated by A.~B.~Zamolodchikov~\cite{Za}, for a
review see~\cite{BSch}.

At least two methods can be used to find the characteristic
integrals. One of them consists in constructing a generating
system of pseudo-differential operators \cite{BiGea, BiGeb, Ba+Fe+,
FORRTW}, and the other one is the Drinfeld--Sokolov highest weight
gauge \cite{DS, Ba+Fe+, FORRTW}. In fact, both of them are based on
the representation of the Toda equations as the zero curvature
condition of some connection on the trivial principal fibre bundle
$\R^2 \times G \to \R^2$.

For the system under consideration the characteristic integrals appear
as the matrix-valued quantities
\begin{eqnarray*}
&& W_1 = {} - \kappa \left[ \Gamma^{(1)-1} \partial_-
\Gamma^{(1)} + \Gamma^{(2)-1} \partial_- \Gamma^{(2)} \right], \\
&& W_2 = {} - \frac{\, \kappa^2}{2} \, \partial_- \left[
\Gamma^{(1)-1} \partial_-\Gamma^{(1)} - \Gamma^{(2)-1}
\partial_-\Gamma^{(2)} \right] \\*
&& \hspace{14em} {} + \kappa^2 \Gamma^{(1)-1} \partial_- \Gamma^{(1)}
\Gamma^{(2)-1} \partial_- \Gamma^{(2)},
\end{eqnarray*}
satisfying relation (\ref{ci1}), and the matrix-valued quantities
\begin{eqnarray*}
&& \bW_1 = {} - \kappa \left[ \partial_+ \Gamma^{(1)} \Gamma^{(1)-1} +
\partial_+ \Gamma^{(2)} \Gamma^{(2)-1}  \right], \\
&& \bW_2 = {} - \frac{\, \kappa^2}{2} \, \partial_+ \left[
\partial_+\Gamma^{(1)} \Gamma^{(1)-1} - \partial_+\Gamma^{(2)}
\Gamma^{(2)-1} \right] \\*
&& \hspace{14em} {} + \kappa^2 \partial_+ \Gamma^{(2)} \Gamma^{(2)-1}
\partial_+ \Gamma^{(1)} \Gamma^{(1)-1} ,
\end{eqnarray*}
satisfying relation (\ref{ci2}). 

Denote the Hamiltonian counterparts of the above described
characteristic integrals by $\cW_1$, $\cW_2$ and $\bcW_1$, $\bcW_2$.
It can be shown that they are the generators of a $W$-algebra with
respect to the Poisson bracket \cite{NiR}.

To find the form of the infinitesimal symmetry transformations
generated by the characteristic integrals we consider first the
quantity
\[ 
\mathcal W_\varepsilon(t) = \int \mathrm dx \, \tr \left[
\varepsilon_1(t, x)
\mathcal W_1 (t, x) + \varepsilon_2(t, x) \mathcal W_2(t, x) \right],
\] 
where $\varepsilon_1$ and $\varepsilon_2$ are arbitrary infinitesimal
matrix-valued functions on $\R^2$ satisfying the relations
\[
\partial_+ \varepsilon_1 = \partial_t \varepsilon_1 
+ \partial_x \varepsilon_1 = 0, 
\qquad 
\partial_+ \varepsilon_2 = \partial_t \varepsilon_2 
+ \partial_x \varepsilon_2 = 0.
\]
The infinitesimal transformations of an observable $F(t)$ generated by
$\cW_\varepsilon(t)$ are given by the relation
\[
\delta F(t) = \{\cW_\varepsilon(t), F(t)\}.
\]
For the mappings $\Gamma^{(1)}$ and $\Gamma^{(2)}$ these
transformations written in the Lagrangian form are \cite{NiR}
\begin{eqnarray}
&& \delta_\varepsilon \Gamma^{(1)} = \Gamma^{(1)} \, \varepsilon_1 -
\kappa \, \Gamma^{(1)} \, \Gamma^{(2)-1} \, \partial_- \Gamma^{(2)}
\, \varepsilon_2 - \frac{\kappa}{2} \, \Gamma^{(1)} \, \partial_-
\varepsilon_2, \label{73} \\
&& \delta_\varepsilon \Gamma^{(2)} = \Gamma^{(2)} \, \varepsilon_1 -
\kappa \, \Gamma^{(2)} \, \varepsilon_2 \, \Gamma^{(1)-1} \,
\partial_- \Gamma^{(1)} + \frac{\kappa}{2} \, \Gamma^{(2)} \,
\partial_- \varepsilon_2. \label{74}
\end{eqnarray}
Similarly, introducing the quantity
\[ 
\bcW_\varepsilon(t) = \int \mathrm dx \, \tr \left[
\bar \varepsilon_1(t, x) \bcW_1 (t, x) + \bar \varepsilon_2(t, x)
\bcW_2(t, x) \right],
\]
where the infinitesimal matrix-valued functions $\bar \varepsilon_1$
and $\bar \varepsilon_2$ satisfy the relations
\[
\partial_- \bar \varepsilon_1 = \partial_t \bar \varepsilon_1 
- \partial_x \bar \varepsilon_1 = 0, 
\qquad 
\partial_- \bar \varepsilon_2 = \partial_t \bar \varepsilon_2 
- \partial_x \bar \varepsilon_2 = 0,
\]
we come to the following expressions for the infinitesimal
transformations~\cite{NiR}
\begin{eqnarray}
&& \delta_{\bar \varepsilon} \Gamma^{(1)} = \bar \varepsilon_1 \,
\Gamma^{(1)} - \kappa \, \bar \varepsilon_2 \, \partial_+
\Gamma^{(2)} \, \Gamma^{(2)-1} \, \Gamma^{(1)}  - \frac{\kappa}{2} \,
\partial_+ \bar \varepsilon_2 \, \Gamma^{(1)}, \label{75} \\
&& \delta_{\bar \varepsilon} \Gamma^{(2)} = \bar \varepsilon_1 \,
\Gamma^{(2)} - \kappa \, \partial_+ \Gamma^{(1)} \, \Gamma^{(1)-1} \,
\bar \varepsilon_2 \, \Gamma^{(2)} + \frac{\kappa}{2} \, \partial_+
\bar \varepsilon_2 \, \Gamma^{(2)}. \label{76}
\end{eqnarray}
One can verify that transformations (\ref{73}), (\ref{74}) and
(\ref{75}), (\ref{76}) are really symmetry transformations for the
Toda system under consideration. Putting $\varepsilon_2 = 0$ and
$\bar \varepsilon_2 = 0$ we obtain infinitesimal version of the
transformations described by relations (\ref{ws1}), (\ref{ws2}).

The conformal transformations (\ref{ct}) are generated by the
Hamiltonian counterparts of the nonzero components of the conformally
improved energy-momentum tensor which are connected with the
characteristic integrals as
\[
T'_{--} = \frac{1}{\kappa} \, \tr \left[ W^2_1 - 2 W_2 \right], \qquad
T'_{++} = \frac{1}{\kappa} \, \tr \left[ \bW^2_1 - 2 \bW_2 \right].
\] 
Note that the characteristic integrals $W_1$, $\bW_1$ and $W_2$,
$\bW_2$ have the conformal spins $1$ and $2$ respectively \cite{NiR}. 

Along the same lines one can investigate also some non-abelian Toda
systems related to the real symplectic Lie group $\mathrm{Sp}_n(\R)$
\cite{NiR}. Actually these systems can be considered as the reduction
of the systems defined above to the case where
\[
\Gamma^{(1)} = (\Gamma^{(2)T})^{-1} = \Gamma,
\]
where $T$ denotes the transposition with respect to the main skew
diagonal. The Toda equations take in this case the form
\begin{equation}
\partial_+(\Gamma^{-1} \partial_- \Gamma) = - (\Gamma^T \Gamma^{-1}).
\label{nale}
\end{equation}
In the case $n=1$ denoting $\Gamma = \exp F$ one comes to the equation
\[
\partial_+ \partial_- F = - \exp(-2F)
\]
that is the well known Liouville equation. Therefore, it is natural to
call equation (\ref{nale}) non-abelian Liouville equation.

\section{Some deliberation}

It would be very interesting to extend our consideration to
non-abelian Toda equations based on $\Z$-gradations different from
ones we have considered here \cite{RSaZu}. For this, one should first
construct the corresponding  characteristic integrals. Of special
interest are higher grading Toda-type systems including matter fields.
One of the serious barriers to be overcome in this way is that such
systems, in general, cannot be treated with the help of local
Lagrangians \cite{GeSa}, and so, the construction of conventional
Hamiltonian descriptions seems to be rather problematic.  
 
It is worth to note that the generators of the $W$-algebras considered
above have conformal spin 1 or 2 only. Nevertheless, we have nonlinear
defining relations. As far as we know it is a new phenomenon in the
theory of $W$-algebras.

Another important direction leads from our work to the problem
of quantization of non-abelian Toda systems. There, one cannot
avoid the questions on quantum counterparts of our $W$-algebras
and on the very meaning of the $W$-symmetries at the quantum 
level.
 
The work of A.V.R. was supported in part by the Russian Foundation
for Basic Research under grant \#01--01--00201 and the INTAS under
grant \#00--00561.

\end{document}